# The Trigger System for the External Target Experiment in CSR of HIRFL

LI Min(李敏)[1,2]   ZHAO Lei(赵雷)[1,2;1)]
LIU Jin-Xin(刘金鑫)[1,2] LU Yi-Ming(鲁一鸣)[1,2] LIU Shu-Bin(刘树彬)[1,2] AN Qi(安琪)[1,2]

1 State Key Laboratory of Particle Detection and Electronics, University of Science and Technology of China, Hefei, 230026, China,

2 Department of Modern Physics, University of Science and Technology of China, Hefei, 230026, China

**Abstract:** A trigger system was designed for the external target experiment in the Cooling Storage Ring (CSR) of the Heavy Ion Research Facility in Lanzhou (HIRFL). Considering that different detectors are scattered in a large area, the trigger system is designed based on a master-slave structure and fiber-based serial data transmission technique. The trigger logic is organized in hierarchies, and flexible reconfiguration of the trigger function is achieved based on command register access or overall field-programmable gate array (FPGA) logic on-line reconfiguration controlled by remote computers. We also conducted tests to confirm the function of the trigger electronics, and the results indicate that this trigger system works well.

## 1. Introduction

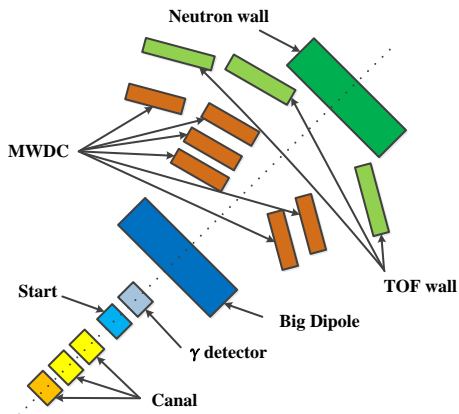

Fig. 1. Architecture of the readout electronics of the CSR external target experiment.

The Cooling Storage Ring (CSR) of the Heavy Ion Research Facility in Lanzhou (HIRFL) is a large-scale comprehensive research center, which consists of a main ring (CSRm), an experiment ring (CSRe), and a radioactive beam line (RIBLL2) to connect the two rings together [1, 2]. At the CSR complex, there are an internal target experiment for hadron physics and an external target experiment for heavy ion collisions [3]. The external target experiment is composed of a start time (T0) detector, a γ detector, a big dipole, one neutron wall, three Time of Flight walls (TOF walls), and six Multi-Wire Drift Chambers (MWDCs), as shown in Fig .1.

The main readout electronics for the detectors in the external target experiment have been designed. For example, the Analog Front End module (AFE) combined with the High-Density Time Digitization Module (HDTDM) were designed for the MWDC. The AFE is used to convert the input charge information to a output pulse width, which is fed to HDTDM for time digitization (the resolution is around 100 ps) [4]. As for TOF walls and neutron wall, the high-resolution Time and Charge Measurement Module (TCMM) was designed, in which charge measurement is achieved based on the TOT method using SFE16 chips [5] and HPTDC chips [6] are used for time digitization, and a time resolution of around 25 ps is achieved in TCMM [7]. A clock module was also designed to generate high precision 40 MHz clock signals for all the front-end measurement modules. To accommodate system-level assembly and extension, PXI 6U [8] standard is employed in the electronics design. As an indispensable part in this readout electronics system, a trigger system is very important for valid data readout and background noise rejection, which is the main work presented in this paper.

*Supported by the National Natural Science Foundation of China (11079003), the Knowledge Innovation Program of the Chinese Academy of Sciences (KJCX2-YW-N27), and the CAS Center for Excellence in Particle Physics (CCEPP).
1) Email: zlei@ustc.edu.cn

In the external target experiment, except MWDCs, all the other detectors will participate in the trigger processing. The front-end measurement modules for different detectors are installed in multiple PXI crates, and besides, these crates are scattered over a large area in the experiment hall. Considering the above situation, the trigger system is organized in three hierarchies, which include the trigger logic blocks in front-end measurement modules, Slave Trigger Modules (STMs), and one Master Trigger Module (MTM), as shown in Fig. 2. The first hierarchy abstracts effective hit information from the detector output signals, and then these hits are further processed in the STM to generate sub-trigger information. MTM collects all the information form STMs and then a global trigger signal is finally generated, which is then distributed to STMs and then to the front-end measurement modules for data readout.

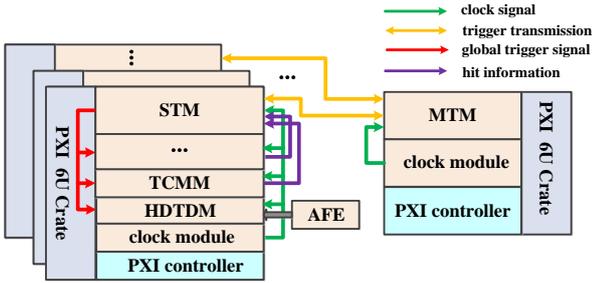

Fig. 2. Architecture of the readout electronics of the CSR external target experiment.

Two main difficulties exist in the design of trigger system:
1) high quality of trigger signal transmission between MTM and STMs considering the long distance between them;
2) Flexible reconfiguration of the trigger logic, since differeent trigger patterns would be needed according to the objects of the nuclear experiments.

To address these issues, we made research in the trigger system design, for example, trigger signal tansmission based on fiber and reconfiguration methods of the trigger logic, which are illustrated in the following sections.

This paper is organized as follows. Section 2 introduces the architecture of the trigger system and the main techniques utilized in it as well as the design of three-hierarchy trigger logic and trigger functionalities. Section 3 describes the laboratory test results, and finally section 4 gives a conclusion.

## 2. Trigger System Design

### 2.1 Architecture of the trigger system

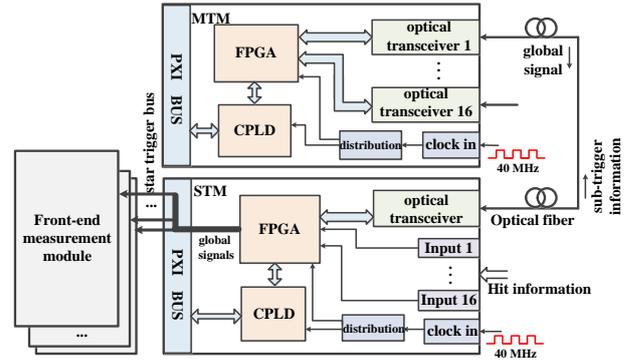

Fig. 3. Architecture of the trigger electronics system.

As aforementioned, in the trigger system, a master-slave structure is designed. Both MTM and STM are based on the PXI 6U standard. The architecture of the trigger electronics system is shown in Fig. 3. In the external target experiment, the front-end measurement modules are distributed in at least 12 PXI crates over a distance up to 100 meters among them. Fibers are employed to guarantee signal transmission quality, since fiber has the advantages of high resistance to EMI (Electro-Magnetic Interference) and capability of isolating the electronic connection among PXI crates. For further extension in future, 16 optical transceivers are integrated in one MTM. As for the STM, it collects the hit information from the front-end measurement modules in the same crate, and generates sub-trigger information which is transmitted to MTM through fibers, as shown in Fig. 3. The MTM then gathers and processes all the sub-trigger information from STMs, and generates a global signal, which is transmitted back to STMs. The STM in each PXI crate further distributes it to different front-end measurement modules through the backplane star trigger bus for valid data readout.

Both in STMs and MTM, the trigger processing and algorithms are built in high density FPGA devices. With the abundant inner connections and logic resources in

FPGAs, reconfigurable trigger electronics can be achieved. The PXI data interface in STM and MTM is implemented in a CPLD with a PCI core (pci_mt32 from the Altera Company) [9] in it. Through this interface and controlling of FPGA configuration based on the PS (Passive Serial) mode, on-line logic modification of these FPGA devices can be achieved, which means the functionalities of these trigger modules can be customized easily according to different physics experiments. Besides, the whole trigger system is also synchronized with the 40 MHz clock, which is shared among all the front-end measurement modules.

In the hardware design, to achieve good system reliability, special care is taken on the signal integrity of both STM and MTM, such as impendence matching, design of complete ground layers in the PCB (Printed Circuit Board), differential transmission of important signals, etc.

**2.2 Fiber-based data transmission**

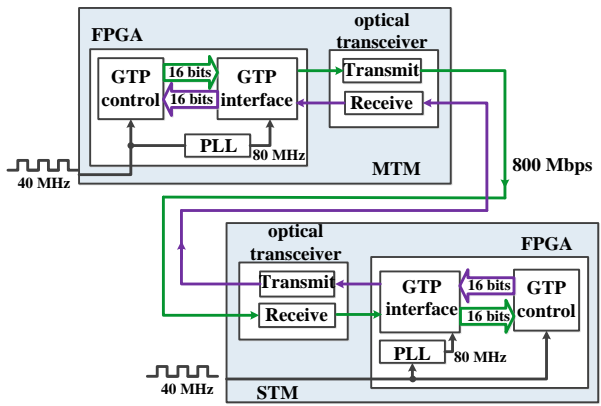

Fig.4. Data transmission between MTM and STM.

In this trigger system, fiber-based transmission method is adopted. We choose FTLF8519P2BNL from Finisar Company as the optical transceiver [10], which features a maximum serial data rate of 2.125 Gbps within a 500 meter distance. To reduce long distance transmission loss, 8B/10B encoding is utilized [11], which greatly decreases bit error probability. Internal GTP interfaces [12] in the FPGA are employed as the bridges between the internal data and the external optical transceivers. Fig. 4 shows the data transmission between the MTM and STM. The 16-bit parallel data are fed to the GTP interface in one module, which are converted to a serial data stream. This data stream is transmitted to the other module, and then the 16-bit wide data are recovered through the GTP interface in it. Bidirectional data transmission can be achieved with this architecture, and high signal quality can be guaranteed over a long distance. Due to clock requirement of GTP interface, the 40 MHz global clock is multiplied to 80 MHz by one Phase Locked Loop (PLL) in the PFGA, and the final serial data transfer rate is 800 Mbps.

After system powering up, the fiber communication is in chaos, so two steps are taken to make it operate normally. Firstly we should initialize the transmitter (TX) and receiver (RX) of GTP interface. Fig. 5 illustrates TX initialization process. This process starts with user command, with which the "INIT_PULSE" signal is generated. The initialization control logic then asserts the "PLLRESET" signal to reset the internal PLL in GTP interface core. Meanwhile, "GTTXRESET" goes high, and it goes low back when "PLLLOCK" is asserted, which indicates completion of PLL reset operation. And then the GTP interface starts to reset the TX datapaths. When the above process is finished, "RESETDONE" turns high, as a flag to indicate that the GTP interface is ready for use. The RX initialization process is just similar.

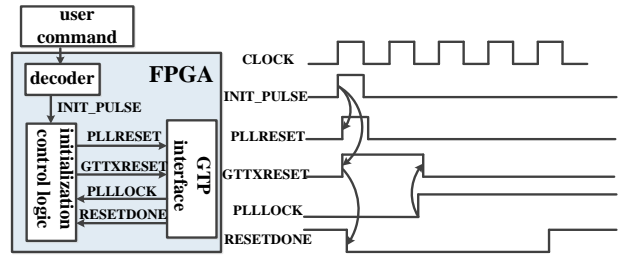

Fig. 5. TX initialization process.

The second step is byte alignment to convert the serial stream to parallel data using GTP. To locate the first bit of the 16-bit parallel data in the serial stream, a special sequence named comma is used. When the receiver detects the comma in the serial stream sent from the transmitter, it moves the comma to a byte boundary so the subsequent data could be well aligned. In this design, code K28.5 is chosen. After these two steps are finished, the bidirectional data communication is established, ready for the transfer of trigger information.

## 2.3 Reconfiguration of trigger function

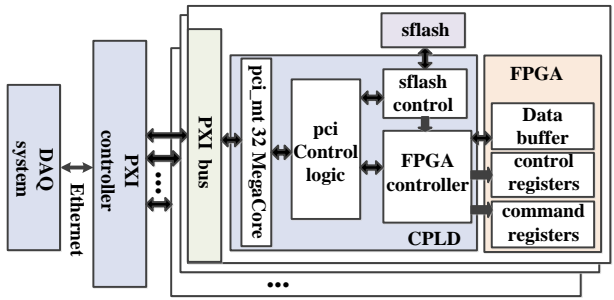

Fig. 6. Block diagram of data transfer and trigger function reconfiguration via PXI interface.

As mentioned above, flexible reconfiguration of trigger logic is preferred due to different requirements of trigger pattern according to the objects of the nuclear experiments. To achieve this, the trigger logic can be modified in two modes: one is partial reconfiguration and the other is on-line modification of the overall trigger function.

As shown in Fig. 6, the PXI controller in each crate receives the user command from DAQ through Ethernet, and then transfers it through the PXI bus to the trigger module in it. In each module, we implement the PXI interface in a CPLD device based on the pci_mt32 MegaCore. The CPLD transfers this command further to the control registers of the trigger logic implemented in FPGA devices. In this mode, since the registers can be modified anytime when needed, flexible partial reconfiguration of the trigger logic can be achieved.

In some special experiments, the trigger pattern could be totally different. In this case, the CPLD receives the configuration data from DAQ, and store them in an external sflash, and then it starts the reconfiguration process of the FPGA with the data. FPGA is set to PS configuration mode, and according to logic resource consumption in STM and MTM, two types of sflashes are employed, M25P128 [13] for MTM and M25P32 [14] for STM, respectively.

## 2.4 Trigger logic design

As mentioned above, the trigger system is organized in hierarchies and the trigger process is executed in three steps. Details are presented in the following sections.

### 2.4.1 Preprocessing logic in front-end measurement modules

In trigger processing of the CSR external target experiment, MWDCs are not engaged. In TOF walls and neutron wall, the signals are read out by the plastic scintillator and a pair of photomultiplier tubes (PMTs) attached to its two ends, and are then transmitted to TCMMs.

When the measurement modules receives hit signals in one event, the particles would strike on different places of the scintillators, which means that the arrival time of two hit signals from each pair of PMT would be different. Therefore, we designed a meantime logic to align these hit signals. In TOF walls and neutron wall, since the delay between two hit signals from a pair of PMT would vary within 17 ns, we expand these hit signals to 25 ns, and generate an effective output hit signal (marked as "meantimer i" in Fig. 7.) through "AND" logic. After counting the hit number within a time period (can be user defined), a special signal will be generated and transmitted to the SMT for the next hierarchy trigger processing, and we use its pulse width to contain the hit number information. Since one TCMM only has 16 input channels, thus the hit number is no more than 8. Besides, the preprocessing logic is also synchronized with the global 40 MHz clock.

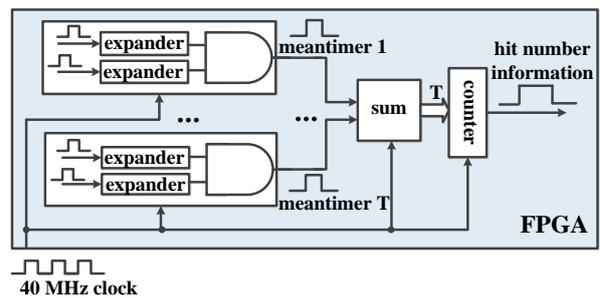

Fig. 7. Preprocessing logic in front-end measurement modules.

### 2.4.2 Trigger processing in STM

Each STM collects hit signals from the front-end measurement modules within the same PXI crate, and performs the trigger processing of the second hierarchy. Different trigger patterns would be required for different experiments. In this paper, we present the logic design for one basic trigger pattern. For example, for the TOF walls and neutron wall, the prerequisite for the trigger processing is that at least one hit is received in the STM. This signal is marked as "sub_trg_flag". According to physics experiment requirement, the total hit number need to be calculated within a certain time range. This is achieved by pulse expansion of input hit signals and "OR" logic to generate the "sub_trg_flag" signal. The expansion time parameter is user defined according to physics requirements. This signal is then used as the start signal for the "Enable Pulse Generator" logic to generate the enable signal used for the "Adder" to sum the hit number of different front-end measurement modules. Since the hit number information is embedded in the pulse width of input hit signals, we recover the information using counters, as shown in Fig. 8. The summed hit number marked as "Tsum" is then synchronized and becomes part of sub-trigger information. And the sub-trigger information is then serialized by the GTP interface and transmitted to the MTM through an optical transceiver. The sub-trigger information is a 16-bit parallel data in a specific format, in which the top 4 bits are the data tag, and the other 12 bits are "Tsum". And to distinguish the sub-trigger information from other code stream, the data tag is set to 4'b0100.

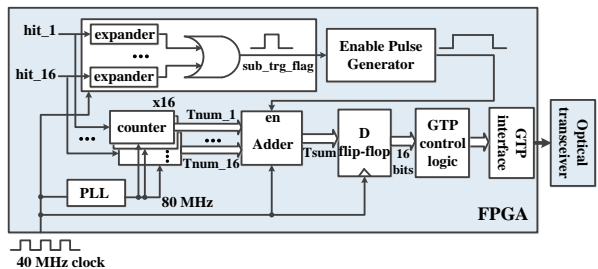

Fig. 8. Trigger processing in FPGA of STM.

Besides, which input hits participate in the above logic can be chosen through configuration according to user command, as discussed in Section 2.3.

### 2.4.3 Trigger processing in MTM

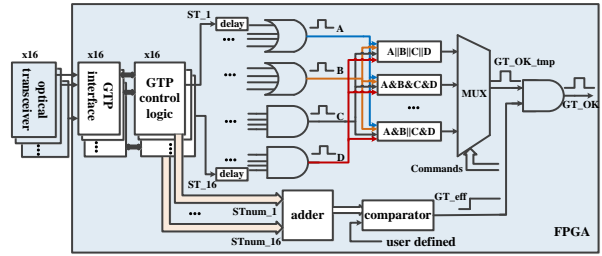

Fig. 9. Block diagram of logic algorithm in FPGA of MTM.

The core trigger logic is built in the FPGA of MTM, where sub-trigger information is gathered together and processed to generate the global trigger signal. The sub-trigger information is received from the GTP interface and then translated to 12-bit hit number data, marked as "STnum_i (i: 1-16)" in Fig. 9. Meanwhile, a flag signal (marked as "ST_i" in Fig. 9) is also generated to indicate that sub-trigger information is received from the corresponding STM (i.e. STM No. i). A global trigger signal is generated on two conditions: one is the total hit number from all the 16 STMs exceeds a predefined value, and the other is that the 16 flag signals (i.e. "ST_1" to "ST_16") concord with a special trigger pattern. As for the first condition, it can be simply implemented by using adder and comparator, so we focus on the logic design for the second one.

Because of different detector types and locations, the arrival time of received sub-trigger information will be different. Thus we should synchronize the 16 flag signals before logic algorithm. As shown in Fig. 9, the delay component is employed to achieve this. And each delay time can be set respectively according to user command.

According to the requirement of the external target experiment in CSR, the 16 flag signals are categorized into four groups, and in each group the combinatory logic is fixed. To guarantee a good flexibility, we aim to change the logic among groups in different experiments. As shown in Fig. 9, MTM switches to a certain logic function according to the user commands from remote PC, so the trigger pattern can be modified on line. If a totally different trigger pattern is required in future experiments, the logic can be updated through FPGA reconfiguration, as mentioned above.

Besides, according to the experiment requirements, the "STnum_i" data are also buffered and transferred to the DAQ system.

Once the global trigger signal "GT_OK" (in Fig. 9) is generated in the MTM, it will be transmitted back to all the STMs, and further fanned out to front-end measurement modules for valid data readout.

## 3 Test Results

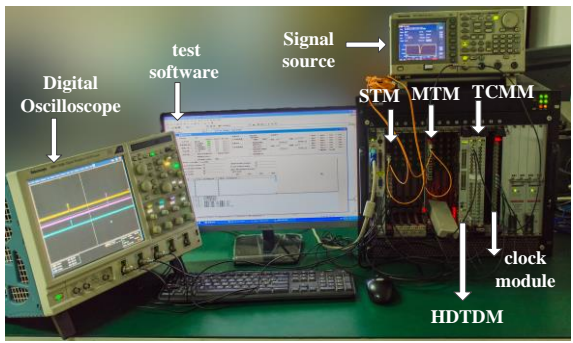

Fig. 10. Laboratory test platform.

To confirm the function of the overall trigger system, we conducted tests in the laboratory. Fig. 10 shows the test platform.

### 3.1 Fiber-based data transmission test

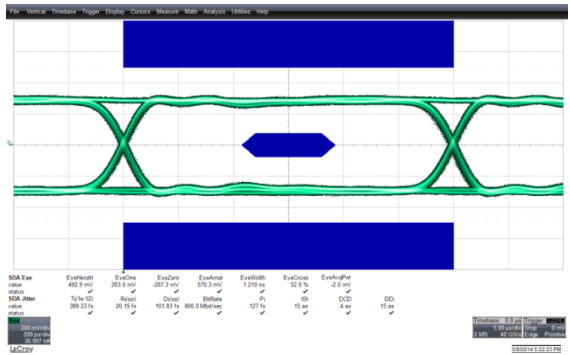

Fig. 11. Eye diagram test result.

To confirm the validity and stability of data communication between the MTM and STMs, we first tested the quality of fiber-based signal transmission. Two test modes were included: eye diagram measurement and BER (Bit Error Rate) test. The eye diagram measurement was conducted with the LeCory WaveRunner 640Zi Digital Oscilloscope [15] and DA300A-AT probe of 4GHz bandwidth. And we plotted the eye diagram with customized template according to GTP interface datasheet. The test results are shown in Fig. 11, which indicates the timing margin and amplitude margin are about 1.21 ns and 490 mV, which are good enough.

PRBS (Pseudo Random Binary Sequence) are usually used as test code stream in The BER test. In this test, we used the special block in the GTP transceiver to generate PRBS-7 pattern. We conducted test on four signal transmission directions between the STM and MTM. As shown in Table. 1, the BER is lower than $10^{-13}$, which indicates a good stability.

Table. 1. BER test results.

| Sending | Receiving | Time(h) | data(bit) | Error | BER |
|---|---|---|---|---|---|
| STM | STM | 24 | $6.912 \times 10^{13}$ | 0 | $<1 \times 10^{-13}$ |
| STM | MTM | 24 | $6.912 \times 10^{13}$ | 0 | $<1 \times 10^{-13}$ |
| MTM | STM | 24 | $6.912 \times 10^{13}$ | 0 | $<1 \times 10^{-13}$ |
| MTM | MTM | 24 | $6.912 \times 10^{13}$ | 0 | $<1 \times 10^{-13}$ |

### 3.2 Trigger logic function test

To evaluate the validity of the trigger electronics, we conducted two types of tests. The first is to observe the important signals generated during the trigger processing using the oscilloscope. In the second test, we verify the trigger function with random hit signals.

First, we used the signal source AFG3252 to generate hit signals, and observe the waveforms of the trigger signals of STM and MTM in a certain trigger pattern. As shown in Fig. 12, the waveforms (from top to bottom) refer to the two hit signals, "sub_trg_flag", and the global trigger signal, which concord with the expected.

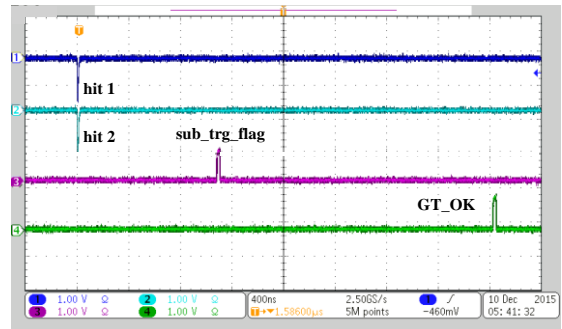

Fig. 12. Waveforms of test signals.

To simulate the situation in actual experiments, we use the linear feedback shift register to generate pseudo random number [16]. Considering that the hit number in each measurement module is from 0 to 8, the random number varies in this range. We used two STMs and one MTM in the test, and in each STM there exist 16 random number generators to simulate the hit signals from the 16 front-end measurement modules within the same crate. Therefore, this test corresponds to the situation in which a total of 32 measurement modules, i.e. 512 channels are included. The test contains 50 runs, and in each run 100 events are generated.

STM, the number of global trigger signals (i.e. valid events) are calculated in each run, and read out to the DAQ system. The results are shown in Fig. 13 (b).

We also built a MATLAB simulation program according to the trigger pattern. The generated pseudo random numbers in the tests are also stored and read out, which are used as inputs for the MATLAB program. The processed results in MATLAB are shown in Fig. 13 (a), which concords well with Fig. 13 (b). The test results indicate that the trigger electronics functions well strictly following the expected trigger pattern.

## 4 Conclusions

A master-slave structure trigger system has been designed for the CSR external target experiment. Utilizing fiber-based signal transmission, long distance communication between master and slave trigger modules is achieved. The overall trigger function is organized in three hierarchies, which makes it easy for further system extension. By implementing trigger logic in FPGA devices, good flexibility can be guaranteed through partial trigger function modification according to user commands or on-line reconfiguration of the FPGA. We also conducted laboratory tests to validate the trigger function, and the results indicate that this trigger system functions well, providing a good technical foundation for the future experiment.

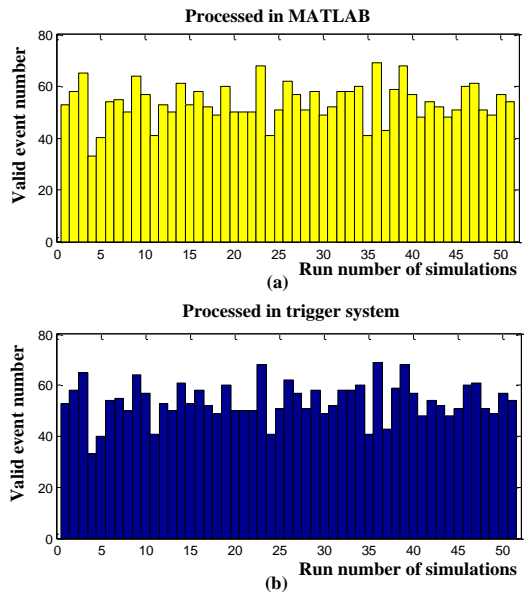

Fig. 13. Valid data ratios calculated in MATLAB and in the trigger system.

For each event, the two STMs process the hit information, and then send sub-trigger information to the MTM for the next level trigger processing. If it is a valid event according to the trigger pattern, a global trigger signal is generated and sent back to the STMs. In the

*The authors would like to thank XU Hu-Shan, SUN Zhi-Yu, YU Yu-Hong, TANG Shu-Wen, and all of the CSR collaborators who helped this paper possible.*

**Reference**

1 Zheng Chuan et al. High Energy Physics and Nuclear Physics, 2007, 31(12): 1177-1180.
2 Xia Jia-Wen et al. Nucl. Instrum. Methods A, 2002, 488: 11–25.
3 ZHAO Lei et al. Nuclear Science and Techniques, 2014, 010401:1-6.
4 Liu Xiao-Hua, The research on the readout electronics system of detectors in CSR, Ph.D. Thesis, University of Science and Technology of China, 2008: 15.
5 E. Delagnes et al. IEEE Trans. Nucl. Sci., 2000, 47(4): 1447-1453.
6 LIU Shu-Bin et al. Nuclear Techniques, 2006, 29(1): 72-76.
7 Zhou Jia-Wen, The research and design on the pre-research readout electronics system of the external experiment in CSR, Ph.D. Thesis, University of Science and Technology of China, 2012.
8 PXI Hardware Specification Revision 2.2, Sept. 2004. http://www.pxisa.org/userfiles/files/Specifications/PXIHWSPEC22.pdf
9 PCI Compiler User Guide, 2005. http://www.altera.com/literature/ug/ug_pci.pdf
10 Finisar, FTLF8519 Data Sheet: http://www.finisar.com/sites/default/files/pdf/FTLF8519P2xNL%20Spec%20RevL.pdf.
11 Widmer A X et al., IBM Journal of Research and Development, 1983, 27: 440-451.

12 Xilinx, 7 Series FPGAs GTP Transceivers. http://www.xilinx.com/support/documentation/user_guides/ug482_7Series_GTP_Transceivers.pdf.
13 ST Microelectronics, M25P128 Datasheet: http://pdf1.alldatasheetcn.com/datasheet-pdf/view/132604/STMICROELECTRONICS/M25P128.html.
14 ST Microelectronics, M25P32 Datasheet: http://pdf1.alldatasheetcn.com/datasheet-pdf/view/22807/STMICROELECTRONICS/M25P32.html.
15 Teledyde LeCroy, WaveRunner 6 Zi Series Datasheet: http://cdn.teledynelecroy.com/files/pdf/waverunner_6_zi_datasheet.pdf.
16 SHU Li-Bao et al. Journal of Circuits and Systems, 2003,1007-0249.